# Validation of FDTD-2D for high-$Q$ resonances description

A.V. Boriskin [1], S.V. Boriskina [2], A. Rolland [3], R. Sauleau [3], A.I. Nosich [1]

[1] *Institute of Radiophysics and Electronics NASU, ul. Acad. Proskury 12, Kharkiv 61085, Ukraine*
a_boriskin@yahoo.com

[2] *Kharkiv National University, Svobody Sqr. 4, Kharkiv 61077, Ukraine*

[3] *IETR, University of Rennes 1, Campus de Beaulieu, bat 11D, 35042 Rennes Cedex, France*

The validation of the FDTD-2D algorithm accuracy for the description of the WGM resonances in circular dielectric resonators made of quartz and silicon is considered. The Mie-type series solution is used as a reference. A loss of accuracy near to high-$Q$ resonances is demonstrated for the near- and far-field characteristics.

**Summary**

FDTD is a powerful method widely applied for the analysis of a wide range of electromagnetic problems [1]. It has gained popularity thanks to simplicity of implementation of numerical algorithms, which are flexible to scatterer geometries and capable of providing informative and reasonably accurate results. The main well-known drawback of the FDTD-3D versions is extremely heavy requirements to computer resources especially in the open-domain and resonance problems. Another drawback, which can easily spoil the analysis of fine resonance effects, is a loss of accuracy near to natural frequencies of studied objects [2].

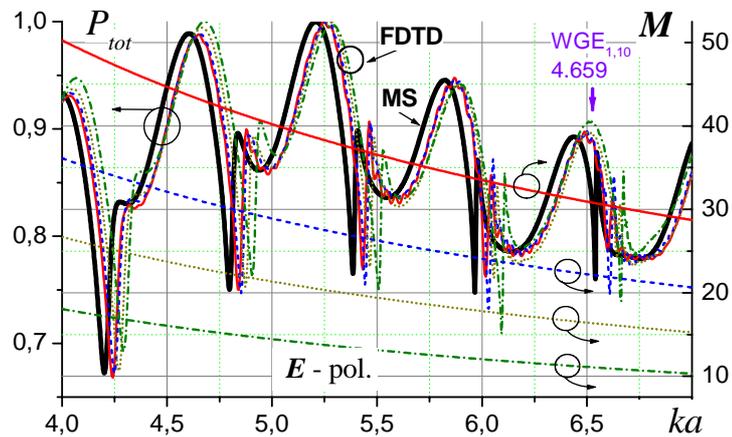

Fig. 1. Normalized far-field characteristics of the line current source illuminating a quartz ($\varepsilon = 4.0$) circular cylinder computed by FDTD and MS, respectively. A family of four curves for FDTD is for different mesh sizes. Mesh size parameter values for each curve are represented by lines of corresponding types associated with the right axis. The mesh size at any frequency point can be estimated as $\lambda_0/M$.

This drawback often escapes attention although it is intrinsic for any 2-D and 3-D version of FDTD and can become a bottleneck in the analysis of small-size dielectric lenses or resonators. This is because in the electromagnetic and optical behavior of such scatterers internal resonances play dominant role - see, e.g., [3-5].

Unfortunately, there are very few papers focused on the validation of the FDTD codes as to reliable characterization of resonance phenomena in dielectric scatterers. As FDTD does not have any in-built criterion of accuracy, it can only be estimated via comparison with more advanced methods providing controllable precision, e.g. Mie-type (MS) series for the circular cylindrical resonator.

The aim of the paper is to assess the accuracy of a standard FDTD-based numerical algorithm in the analysis of the near and far fields of dielectric cylinders. To achieve the goal, we consider circular dielectric resonators excited by the line current sources and demonstrate the failure of FDTD algorithm to characterize high-$Q$ whispering-gallery-mode (WGM) resonances. We do this by comparison with the exact Mie-series solutions (Fig. 1). The near-field and far-field characteristics are studied for resonators of various sizes, made of quartz and silicon. Both *E*- and *H*-polarizations are considered. The computational error, i.e. a resonance frequency shift observed for the FDTD solution, is analyzed and compared for several WGMs having different *Q*-factors.

More details of the current study and a relevant one dedicated to assessment of FDTD accuracy in characterization of the half-bowtie resonances in hemielliptic dielectric lenses can be found in Refs. [6] and [7], respectively.

# Assessment of FDTD-2D accuracy in high-*Q* resonances description

**Sammary**

The objective of the study is to assess the accuracy of a standard FDTD code in description of high-*Q* resonances such as whispering-gallery-mode (WGM) ones excited in circular dielectric resonators.

To achieve the goal, we consider circular resonators excited by line currents and extract the data on FDTD accuracy from comparisons between the FDTD-solution and the exact one obtained by the Mie-series (MS) solutions.

The near-field and far-field characteristics are studied for resonators of various sizes, made of quartz and silicon. Both the *E* - and the *H* - polarizations are considered. The computational error, i.e. a resonance frequency shift observed for the FDTD solution, is analyzed and compared for several WGMs having different *Q*-factors.

**Geometry of the problem**

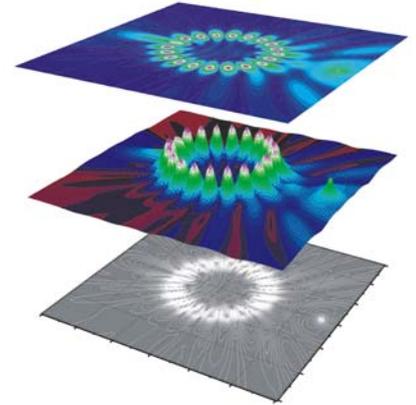

Fig. 1. The near-field map of the $WGE_{1,10}$ resonance excited within a circular resonator made of quartz and illuminated by a cylindrical wave radiated by an *E* - polarized line current.

## Results

**Quartz Circular Resonator Analysis: FDTD-2D vs. MIE Series**

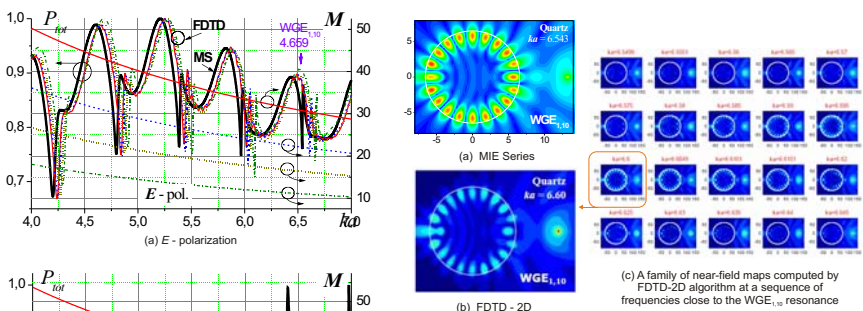

(a) *E* - polarization

(b) *H* - polarization

Fig. 2. Normalized far-field radiated power of a line current source illuminating a quartz circular resonator ($\varepsilon$ = 4.0) of radius *a*, computed by FDTD and MS, respectively. A family of four curves for FDTD is for different mesh sizes. Mesh size parameter values for each curve are represented by lines of the corresponding types associated with the right axis. The mesh size at any frequency point can be estimated as $\lambda_0/M$.

(a) MIE Series

(b) FDTD - 2D

(c) A family of near-field maps computed by FDTD-2D algorithm at a sequence of frequencies close to the $WGE_{1,10}$ resonance

Fig. 3. Normalized near-field maps of quartz ($\varepsilon$ = 4.0) circular resonators excited a line *E*-polarized current located at $x_s$ = 2*a*. The corresponding resonance is marked by an arrow in Fig. 2a.

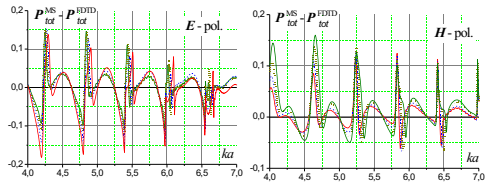

Fig. 4. The computational error of the FDTD-2D solution vs. the normalized frequency computed as a difference between curves presented in Fig. 2, i.e. the MS-curve and the FDTD-curves obtained with different meshes.

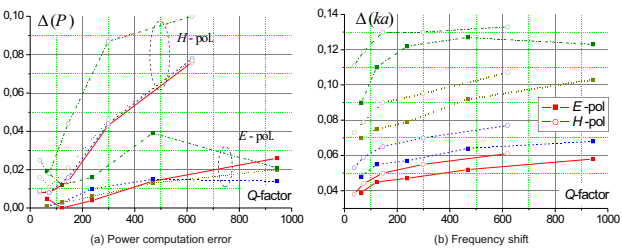

(a) Power computation error

(b) Frequency shift

Fig. 5. Computation errors of the FDTD-2D solutions of the WGM resonances observed in Fig. 2 extracted from comparison between FTDT and MS solutions. A family of four curves is for different mesh sizes.

Q-factors are obtained by solving the corresponding eigen-value problems.

**Silicon Circular Resonator Analysis: FDTD-2D vs. MIE Series**

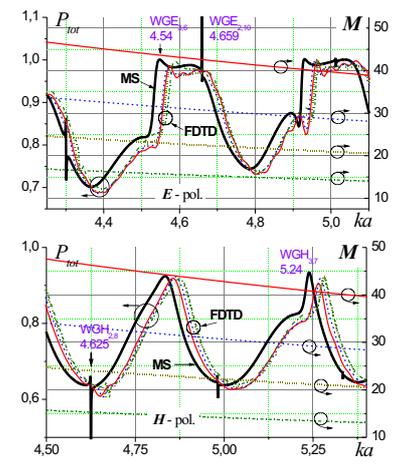

Fig. 6. The same as in Fig. 2 for silicon cylindrical resonator ($\varepsilon$ = 11.7).

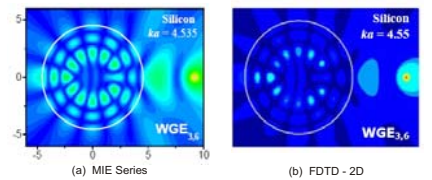

(a) MIE Series       (b) FDTD - 2D

Fig. 7. The same as in Fig. 3 for silicon cylindrical resonator ($\varepsilon$ = 11.7).

## Conclusions

- FDTD-2D- and MS-based numerical algorithms have been implemented and applied for the near- and far-field analysis of circular resonators made of quartz and silicon.
- It has been shown that although FDTD demonstrates sufficiently high accuracy when out of high-*Q* resonances, near to such resonances the error in computations, *i.e. a frequency shift and errors of the total radiated power estimation,* can be unacceptably high. Both errors grow rapidly with increasing of the resonances *Q*-factors. Finer meshing only reduces these errors to the level determined apparently by the type of absorbing boundary conditions used, shape and size of computational window, and other details of the FDTD code (like the staircase approximation). Therefore these widely spread tools of numerical simulation should be used with a caution when applied for the analysis of dielectric objects such as small and medium size resonators or lenses of specific shapes especially made of high-index materials.

**Acknowledgment**

This work was supported by the joint projects of the National Academy of Sciences, Ukraine and Centre National de la Recherche Scientifique, France

The first author was also supported by the National Foundation for Fundamental Research, Ukraine via a Young Scientist Research Grant of the President of Ukraine.

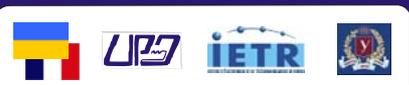

**Artem V. Boriskin[1], Svetlana V. Boriskina[3], Anthony Rolland[2], Ronan Sauleau[2], Alexander I. Nosich[1]**
[1] Institute of Radiophysics and Electronics NAS Ukraine, Kharkov, 61085, Ukraine
[2] Institute of Electronics and Telecommunications, Universite de Rennes 1, Campus de Beaulieu, Rennes, 35042, France
[3] Kharkov National University, Kharkov, 61004, Ukraine